# Road to Serenity: Individual Variations in the Efficacy of Unobtrusive Respiratory Guidance for Driving Stress Regulation

A.J. Béquet[1*], C. Jallais[1], J. Quick[1], D. Ndiaye[1], A.R. Hidalgo-Muñoz[2]

1. *Laboratory Ergonomics and Cognitive Sciences applied to Transport, TS2-LESCOT Univ Gustave Eiffel, IFSTTAR, Univ Lyon, F-69675, Lyon, France*
2. *Institute of Neuroscience of Castilla y León (INCYL), University of Salamanca, Spain*

* Corresponding author : Adolphe J. Béquet

   Email : adolphe.bequet@univ-eiffel.fr

   25, avenue François Mitterrand

   69675 Bron Cedex – France

**Abstract**

Stress impacts driving-related cognitive functions like attention and decision-making, and may arise in automated vehicles due to non-driving tasks. Unobtrusive relaxation techniques are needed to regulate stress without distracting from driving. Tactile wearables have shown efficacy in stress regulation through respiratory guidance, but individual variations may affect their efficacy.

This study assessed slow-breathing tactile guidance under different stress levels on 85 participants. Physiological, behavioral and subjective data were collected. The influence of individual variations (e.g., driving habits and behavior, personality) using logistic regression analysis was explored.

Participants could follow the guidance and adjust breathing while driving, but subjective efficacy depended on individual variations linked to different efficiency in using the technique, in relation with its attentional cost. An influence of factors linked to the evaluation of context criticality was also found.

The results suggest that considering individual and contextual variations is crucial in designing and using such techniques in demanding driving contexts. In this line some design recommendations and insights for further studies are provided.

1. Introduction

**1.1 Stress and Driving**

Driving is a complex activity taking place in a dynamic environment (Neboit, 1982) requiring constant situational awareness (Endsley, 1995). Managing this activity constantly involves complex cognitive processes such as executive functions and attentional mechanisms (Ranchet et al., 2012). Impairments of the processes during driving can therefore lead to uncomfortable and critical situations endangering not only the safety of the driver, but also of other road users.

Research evidenced how emotional processes can influence normal cognitive functioning (Ledoux, 1989). Specifically, stress has been shown to have significant impacts on attentional and decision-making processes. With stress, the altered functioning of key brain regions can lead to manifestations such as perseveration, attentional tunneling toward stressors, and automatic responses (Dehais et al., 2019; Eysenck et al., 2007).

Stress is frequently experienced in driving (Taylor, 2018). Driver stress can be defined as "the set of responses associated with the perception and evaluation of driving as being demanding or dangerous relative to the individual's driving capabilities" (Gulian et al., 1989, p586). According to Matthews' transactional driver stress model, stress can thus emerge from the complex interaction of environmental stressors (eg. Traffic, weather) with individual factors (eg. Personality, appetency for driving), and can lead to performance (eg. Loss of attention) and subjective (eg. Worry) outcomes (Matthews, 2002). In addition, life-related stressors can also influence the generation of stress in driving (Rowden et al., 2011).

In the recent years, technological advance linked to driving automation could seem to be a solution to mitigate stress in driving, as the driver would no be involved in the driving task. However, stress might not only be favored in this context by the possibility to realize certain non-driving related tasks (NDRT) during the phases of automated driving (Naujoks et al., 2018), but could also impact the manual takeovers that are still required in current level of automation (SAE, 2021).

Given its deleterious impacts during driving, stress regulation is desirable in such contexts. However, due to the level of the cognitive and attentional demand elicited by driving, such regulation has to be unobtrusive to avoid further perturbation of the driving task.

**1.2 Unobtrusive stress regulation**

Recent years have seen major developments regarding emotional regulation in terms of conceptualization and implementation. Notably, Gross (2015) proposed that emotion regulation can be antecedent-focused, i.e. oriented on the situation eliciting an emotion, or response focused, i.e. oriented on the individual response elicited by the emotion. In this line, Gross's process model of emotion regulation outlines five strategies related to the dynamics of the emotion: situation selection, situation modification, attentional deployment, cognitive change, and response modulation. Situation selection and modification refer to impacting the situation itself by either avoiding or modifying it to change its emotional aspect. Attentional deployment and cognitive change are linked to the relationship between the individuals and their environment: these strategies consist in either switching their attention toward positive or neutral aspects of the situation, or changing their appraisal. Finally, response modulation proposes to influence the emotion itself, acting on the behavioral, subjective, or physiological responses elicited. These strategies have been implemented in daily life, through the use of various technologies. These include smartphones applications (Drissi et al., 2020), ambient lights and displays (Yu et al., 2018), personalized music (Hu et al., 2015) or wristbands (Costa et al., 2017).

Recent works showed that such techniques could be implemented in demanding contexts, where emotional regulation requires to be unobtrusive (eg. office work, Van Der Zwaag et al., 2013; or public speech Costa et al., 2017). In driving, unobtrusive stress regulation remains an emergent topic. A literature review conducted by Béquet et al., (2020) illustrated the scientific relevancy and technical feasibility of unobtrusive regulation methods matching the model of emotion regulation of Gross (2015). Amongst these methods, physiological modulation through the use of a respiratory guidance is particularly promising for several reasons. Firstly, stress regulation based on slow breathing has been shown to be effective and to present strong effects, given its influence on physiological and cognitive processes (Allen et al., 2022; Zaccaro et al., 2018). Nonetheless, In driving, the use of respiratory guidance has been tested in few studies. In a original study, Lee et al., (2021) used auditory and wind modality to influence the breathing rate of participants. Their results showed that the participants could engage with the breathing intervention after a brief training on the use of the intervention. They also showed that the efficiency of the intervention was linked to the complexity of the driving task, the intervention being more helpful when the car was driving autonomously. Balters et al. (2020) tested tactile respiratory guidance embedded in the driver's seat, and showed that it was effective to reduce the subjective stress level. Zepf et al. (2021) tested a similar technique, showing that its physiological efficacy was increased when an audio guide was also presented. A recent study, from Zhou et al. (2024) investigated the impact of auditory comments together with a respiratory guidance using an apple watch to deliver stimulations at a fixed rate of 7 vibrations per minute. Despite an effect of the guidance to effectively regulate the breathing rate and the physiological metrics associated with stress, their study did not show an effect of the intervention on the subjective levels of stress. Their study suggests an influence of the complexity of the driving task on the efficiency of the physiological stress regulation. While the mentionned studies did not find an impact of the technique on the safety of driving, Zepf et al. (2020) showed that a conscious auditory guidance might increase driving mistakes.

In demanding contexts, the use of single tactile modality to deliver physiological stress regulation intervention has been showed to be relevant, as this modality limits disturbance with the ongoing task (Choi & Ishii, 2020). However, prior research emphasized the importance of personalizing such interventions. Notably, in the work of Umair et al., (2021), several tactile patterns for affect regulation were tested, with the goal to simulate an heartbeat. While focusing on the cardiac modality, their work shows the importance of using slow personalized paced rhythms when mimicking physiological rhythms, as a slow pace might elicit an entrainment effect and be associated with lower anxiety. Specifically, in driving, the importance of the consideration of user profile when using conscious interventions has been proposed by (Zepf et al., 2020). Less experienced drivers and participants with higher levels of neuroticism made more infractions when following the guidance. In the same line, they showed that a higher driving experience, as well as higher extraversion and openness levels were associated with a better subjective perception of the breathing intervention. The influence of personality traits has been suggested in a variety of stress regulation studies. Miri et al. (2020) suggested that higher extraversion levels are associated with reduced distraction from respiratory guidance, leading to enhanced efficacy in anxiety reduction. In the same line, Xu et al. (2021) demonstrated that interoceptive accuracy influenced the physiological efficacy of a tactile stimulation mimicking cardiac beats, suggesting an important contribution of the easiness to attend to physiological activity. Using a comparable technique, Béquet et al., (2022a) explored the influence of extraversion and neuroticism and found that these personality traits impacted subjective stress reduction.

Taken together, the previous studies indicates that in a demanding context such as driving, participants can follow a breathing guidance. However, the modality used to deliver the intervention seems important to consider in demanding contexts such as driving, as previous studies hinted how the use of a counscous regulation based on auditory modality might increase the cognitive load and alter the efficiency of the intervention. Tactile modality seems to be relevant as it exploits a modality less used in driving than audition or vision. However, despite positive impacts of tactile breathing guidance in terms of subjective regulation and driving safety, the factors determining these outcomes need to be further explored. Based on the previous findings, we suggest that the efficiency of the regulation could be linked to the attentional cost of the guidance itself. This factor may be related to complex interactions between individual caracteristics liked to personality, interoceptive abilities and driving experience, as well as to the perceived difficulty of the driving task.

### 1.3 Objectives of the study

As individual variations can impact the efficacy of tactile respiratory regulation during driving, gaining a deeper understanding of their role becomes crucial. In this context, the present study pursued two primary objectives:

-To test the efficacy of a tactile respiratory guidance in unobtrusively reduce the respiratory rate in driving. Regarding this objective, we hypothesized that participants would be able to follow the respiratory guidance while driving safely.

-To identify individual variations that may influence the subjective efficacy of the proposed regulation. Regarding this objective, we hypothesized that subjective stress reduction should vary amongst participants due to individual variations linked to personality, interoceptive abilities, driving styles, and context-specific perceptions.

Regarding the implementation of the regulation, we chose a wristband for various reasons including: direct skin contact regardless of the driver's posture, portability and familiarity for participants, and its existence in certain commercial smartwatches (eg. Apple Watch[1]). Given the increasing prevalence of driving automation contexts, an automated driving scenario combined with intermittent manual control was employed. The use of this context also allows stress induction through a controlled non-driving related tasks (NDRT) adaptive to the driver. Part of this work has been presented in Béquet et al., (2022b, 2023).

## 2. Methods

### 2.1 participants

Eighty-eight adult participants, recruited via social networks, took part in the study. They all had normal hearing and normal or corrected-to-normal vision. Exclusion criteria included declared cardiovascular, respiratory, or neurological diseases, and medical treatments affecting vigilance. None had previously participated in a study using the same materials. Five participants who experienced discomfort from the driving simulator, known as motion sickness (Brooks et al., 2010), were excluded. The final sample comprised 83 healthy French participants, aged 19 to 71 years (M = 39.4, SD = 16.02), with 82 right-handed individuals and 43 men. This sample size was sufficient to ensure statistical power, as illustrated by previous studies (eg. Miri et al., 2020). All held a driving license for at least 2 years and declared to drive frequently. The study adhered to the Declaration of Helsinki and obtained ethical approval from the local committee at University Gustave Eiffel. Participants provided written consent and received €50 as compensation.

---

[1] https://support.apple.com/en-ca/guide/watch/apd371dfe3d7/watchos

### 2.2 Task

Participants drove a driving simulator following 3 phases. First, the simulator was in automated mode to allow the participant to realize a NDRT inducing stress (S) or not (NS). Second, a takeover request instructed the participant to resume the manual control of the vehicle. Third, participants were driving the simulator in manual mode while the respiratory guidance was activated (RG) or not (NRG). These phases are presented in Figure 1 below.

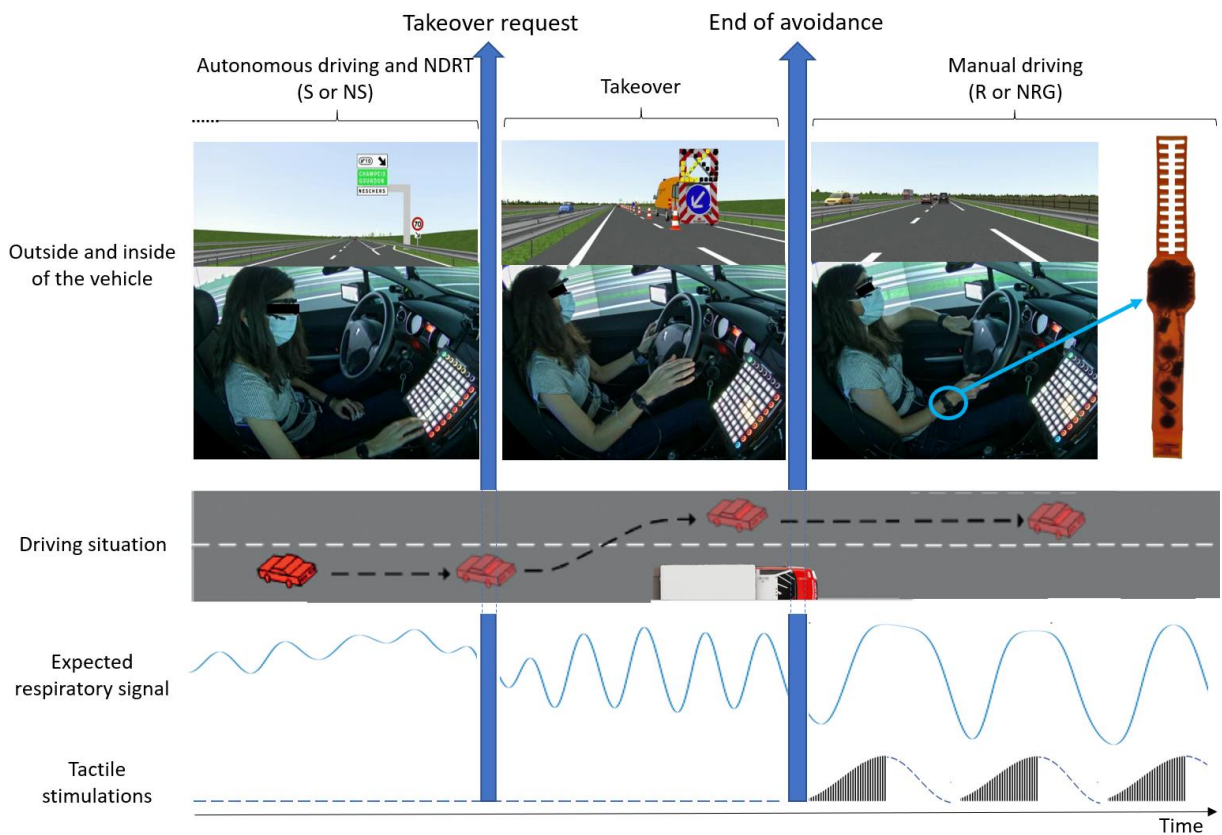

*Figure 1: Illustration of the task and of the disposition of the material. We show the NDRT mode (S: Stress; NS: NoStress), the activation of the respiratory guidance (RG: Respiratory Guidance; NRG: No Respiratory Guidance) and the expected respiratory signal in the conditions where the guidance is activated.*

### 2.2.1 Stressful non-driving related task

The non-driving related task consisted of a memory game validated in a previous study for stress induction (Béquet et al., 2022a).

The memorization game was implemented on a Novation Launchpad MKII, an 8×8 LED button grid with additional LED rows on both sides, offering ergonomic and attractive features (Figure 2). Programmed in Python using the "launchpad_py" toolbox, the game involved memorizing and reproducing green light patterns displayed on the grid. Participants used only their right hand to replicate the patterns, validated their responses by pressing a green button, and received performance feedback via blue circles (success) or red crosses (failure). Difficulty levels varied based on pattern length and adapted to participants' performance, ensuring engagement and manageable challenge. The game had non-stressful and stressful modes inspired by the Montreal Imaging Stress Task (Dedovic et al., 2005). Specifically, in the stressful condition, stress was manipulated through temporal, performance, and social pressures using additional LED rows. Temporal pressure involved a time constraint gauge that decreased based on previous performance. Performance pressure displayed the participant's score with an objective to reach. Social pressure was induced by instructions given by the experimenter: participants were told that the score displayed on the gauge was reflecting their score relative to the other participants of the study and it was specified that if the participant failed to reach the objective, his data would not be considered in the data analysis due to too much deviation from the other participants.

The performance and time gauge were manipulated in real time to ensure that the participant was always struggling to react the objective.

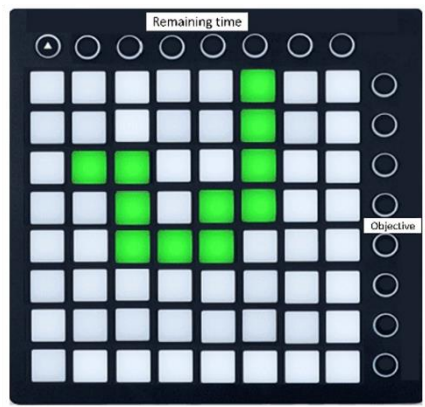

*Figure 2 : Presentation of the memorization game used for stress induction. A light pattern can be seen being displayed, as well as the two additional LED rows used to induce stress.*

### 2.2.2 Driving simulator and scenario

A Peugeot 308 cabin at University Gustave Eiffel was used, located on the Lyon-Bron campus, as a realistic simulator for the study. This immersive cabin featured 10 rear-projection screens, offering a nearly 360° horizontal projection field and a vertical field of 45°. The steering wheel also had a force feedback system.

Participants drove on a 2×2 lane highway. First, the car was in autonomous mode (AD) for 7 to 9 minutes, at a speed of 110km/h while participants performed the NDRT. Due to a stationary vehicle in the lane, an audible alarm instructed participants to resume manual control (TO) of the vehicle with a time budget of 11 s. Participants then manually drove (MD) the vehicle for approximately 5 minutes. A training scenario of less than 5 minutes was built and followed these same steps. Technical details for the road base used are available in Ndiaye et al. (2021).

## 2.3 Respiratory guidance

### 2.3.1 Wristband design

Contrary to previous studies using commercial smartwatches, we chose to design our own wristband in order to control key designs aspects. A 3D-printed wristband was designed by the research team, inspired by the design from Costa et al., (2016). We also considered participant's feedback regarding the design of a wristband used in a previous study Béquet et al., (2022a), were the participants indicated that "*a one piece wristband would be better looking*". The experimenter's feedback regarding the practical use of the wristband in the experimentation was also considered to improve the strength and ease of placement of the material used. This tool consists of 3 coin vibrations motors (Precision Microdrives© 310-122) connected to a bluetooth micro-controller board (DFRobot Beetle BLE©), the whole being encapsulated in an envelope made in TPU 1.75 filament (NinjaFlex©). The motors are 10mm in diameter and 3.4mm thick, have a maximum amplitude of 1.9 G each, and a 14k rpm vibration speed. Three motors were placed to cover the inner side of the dominant (right) wrist, to ensure a good perception of the stimulations, as previous studies points an increased sensitivity to vibrotactile stimulations in this area (Ævarsson et al., 2022). A representation of the wristband is displayed in Figure 3.

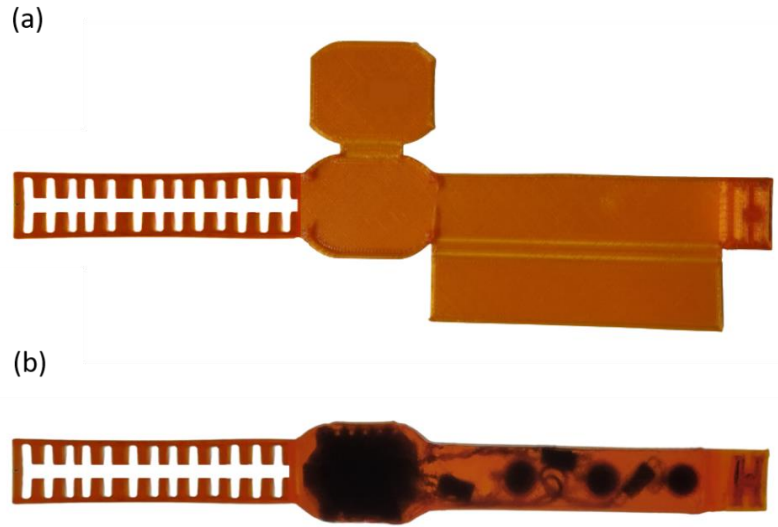

*Figure 3: Representation of the wristband with, (a) 3D printed envelope, (b) transparent view of the components.*

### 2.3.2 Tactile stimulations

The Arduino board was reprogrammed through the Arduino development environment (version 1.8.12) to deliver a certain vibratory pattern upon receiving a trigger. For this study, the vibratory pattern consisted of slight vibrations starting at an amplitude of 10% of the maximum vibration amplitude and gradually increasing to reach 35% of the maximum amplitude. The minimum amplitude was determined based on 5 pretests and corresponded to the threshold of vibration perception by pretested participants. Once the maximum amplitude was reached, the vibrations stopped, following a half-Gaussian curve of vibration intensity, as illustrated in Figure 4. The participant was instructed to inhale during the vibration phases and exhale during the stop phases and performed a training session of about 5 minutes).

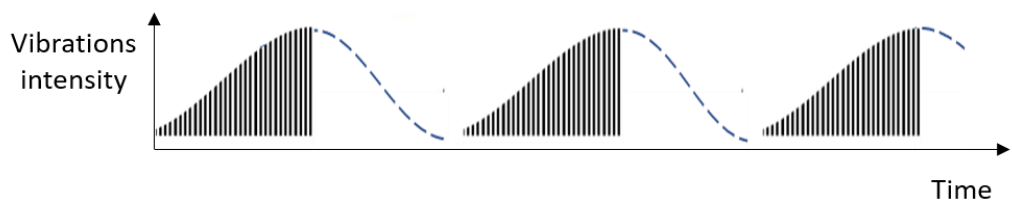

*Figure 4: Illustration of the vibratory patterns delivered at the level of the wristband. The vertical lines represent the vibrations and the dotted lines represent the pauses between vibrations.*

We chose to deliver a target rhythm corresponding to a 30% decrease from the participant's resting respiratory rate (RRR) in breath per minute (BPM) measured during a resting phase, as recommended by previous studies (Choi et al., 2022; Balters et al., 2020). Thus each guided breath (defined as an inspiration and a expiration instruction) was calculated as:

$$Length\ (s) = \frac{60}{(0.7 \times RRR)}$$

In our sample, the length of each guided breath varied from 4 seconds (max RRR of 21 BPM leading to a guided rate of 15 BPM) to 12 seconds (min RRR of 7 breath per minute leading to a guided rate of 5 BPM), with a mean duration of 6.2 seconds (mean RRR of 13.8 BPM leading to a mean guided rate of 9.7 BPM).

### 2.4 Data collection & processing

Physiological and driving data was collected and synchronized using the software RTMaps (version 4.6.0), and processed using dedicated MATLAB (version R2020b) scripts.

#### 2.4.1 Individual traits and feelings

Regarding individual traits, a variety of questionnaires were administered. We used the Big Five Inventory (John et al., 2012) to measure personality, as this model is consensual for personality assessment and enables its subdivision into five major dimensions: extraversion, neuroticism, agreeableness, conscientiousness, and openness. The Three-domain Interoceptive Sensations Questionnaire (THISQ, Vlemincx et al., 2021) was used to assess an individual's attention to physiological states in everyday life. It measures three dimensions across different interoceptive domains: increased cardiorespiratory activity, decreased cardiorespiratory activity, and gastrointestinal activity. Regarding driving, we used the 23-items version of the Driver Behavior Questionnaire (DBQ, Reason et al., 1990), assessing driving behaviors on 6 dimensions (Guého et al., 2014), including dangerous errors, inattention errors, inexperience errors, ordinary or aggressive violations, as well as positive (prosocial) behaviors towards other road users. This questionnaire is standardized and widespread used as a user-friendly tool for measuring driving styles. We also gathered information about the participants' physical fitness level on the day of the assessment and their frequency of driving to ensure their ability to participate in the experiment. Additionally, demographic questionnaires (e.g., gender, age, education level) were conducted prior to the experiment, along with inquiries about the participants' knowledge and use of guided breathing techniques.

Participants' emotional experiences were assessed using a Geneva Emotion Wheel (GEW, Scherer, 2005) consisting of 10 items on a 6-point Likert scale. We included items from the Short Stress State

Questionnaire (Helton, 2004) on the wheel that reflected negative feelings associated with stress (e.g. *frustration*, *social worry*). This was done to facilitate quick and intuitive completion of the questionnaire for the participants.

After the experiment, we obtained participants' feedback about the wristband usability using a debrief questionnaire based on Miri et al. (2020). The questionnaire utilized a scale ranging from 0 to 100. Participants were asked to rate the difficulty in distinguishing between inspiration and expiration phases, the challenge of synchronizing with the delivered rate, and the level of distraction caused by the wristband during the driving task.

### 2.4.2 Behavioral parameters

The data from the simulator was collected at an average frequency of 80 Hz. From these data, we calculated indicators linked to the lateral and longitudinal control of the vehicle in each phase of driving (takeover & manual driving) following SAE's technical guidelines (SAE, 2015). Notably, we calculated the steering wheel reversal rate (SWRR) as the number per minute of steering wheel reversals exceeding 6 degrees of angle. This metric reflects steering control instability (Ameyoe et al., 2015), and is associated with distraction regarding driving (Kountouriotis et al., 2016). We also evaluated the vehicle's speed as a metric to assess the participant's evaluation of the criticality associated with the takeover (Deniel, 2019). Furthermore, we calculated the proportion of participants still performing an action on the NDRT after the takeover alert was sent (NDRT perseverance).

### 2.4.3 Physiological parameters

Physiological data were measured using a BIOPAC MP150© device, coupled with the Bionomadix© wireless system, and processed using a dedicated MATLAB script. The cardiac signal was recorded via electrocardiogram (ECG) at a sampling rate of 1 kHz (hardware low-pass filtered at 35 Hz) using electrodes placed on the right clavicle, left floating ribs, and right hip. The respiratory signal (RESP) was collected at a sampling rate of 1 kHz (hardware low-pass filtered at 10 Hz) using a thoracic belt with a transducer around the participant's chest. Electrodermal activity (EDA) was recorded at a sampling rate of 1 kHz (hardware low-pass filtered at 10 Hz) using electrodes placed on the distal phalanx of fingers II and III of the participant's non-dominant hand, as recommended by Boucsein (2012).

For data processing, the heart rate (HR) and its variability (RMSSD) were extracted using automated detection of R-peaks with the findpeaks function in MATLAB. The respiratory rate (RR) in cycles per minute was obtained from respiratory peaks. The respiratory sigh rate per minute was calculated using respiratory amplitudes greater than twice the average amplitude of breaths (Vlemincx et al., 2013).

Using the software Ledalab (version 3.4.6, Benedek & Kaernbach, 2010) we extracted the number (SCRs) and amplitude of electrodermal responses greater than 0.1μS (Boucsein, 2012). The values of these parameters were normalized using the baseline. As in Béquet et al. (2022a), we calculated the normalized parameter P*norm* considering the relative difference of the non-normalized parameter P*x* with the baseline parameter P*base* as :

$$P_{norm} = \frac{P_x - P_{base}}{P_{base}}$$

### 2.5 Overview of the experimental procedure and design

Figure 5 gives an overview of the conditions and the experimental procedure.

Regarding the training section, there were 3 phases, each of them lasting for about 5 minutes. First, participants were trained to drive the simulator and react to the takeover request, they completed the questionnaires after this session, and their performance was monitored by the experimenter who also verbally checked with them that their filling of questionnaire was coherent with their state. Second, they had to play with the game designed to induce stress, to understand its rules and goal, then again, their verbal feedback was collected to ensure a good comprehension of the mechanics of the game. Finally, the participant wore the wristband and were instructed to follow the personalized rate that was delivered for at least five minutes, as recommended by Lee et al., 2021. A visualization of both the respiratory signal and the tactile signal on a dedicated screen was monitored by the experimenter to ensure that the participant could follow the guidance. If a participant had a difficulty to adjust their rate, the experimenter provided a feedback, and the training lasted until a good match between the two signals was evident on the screen.

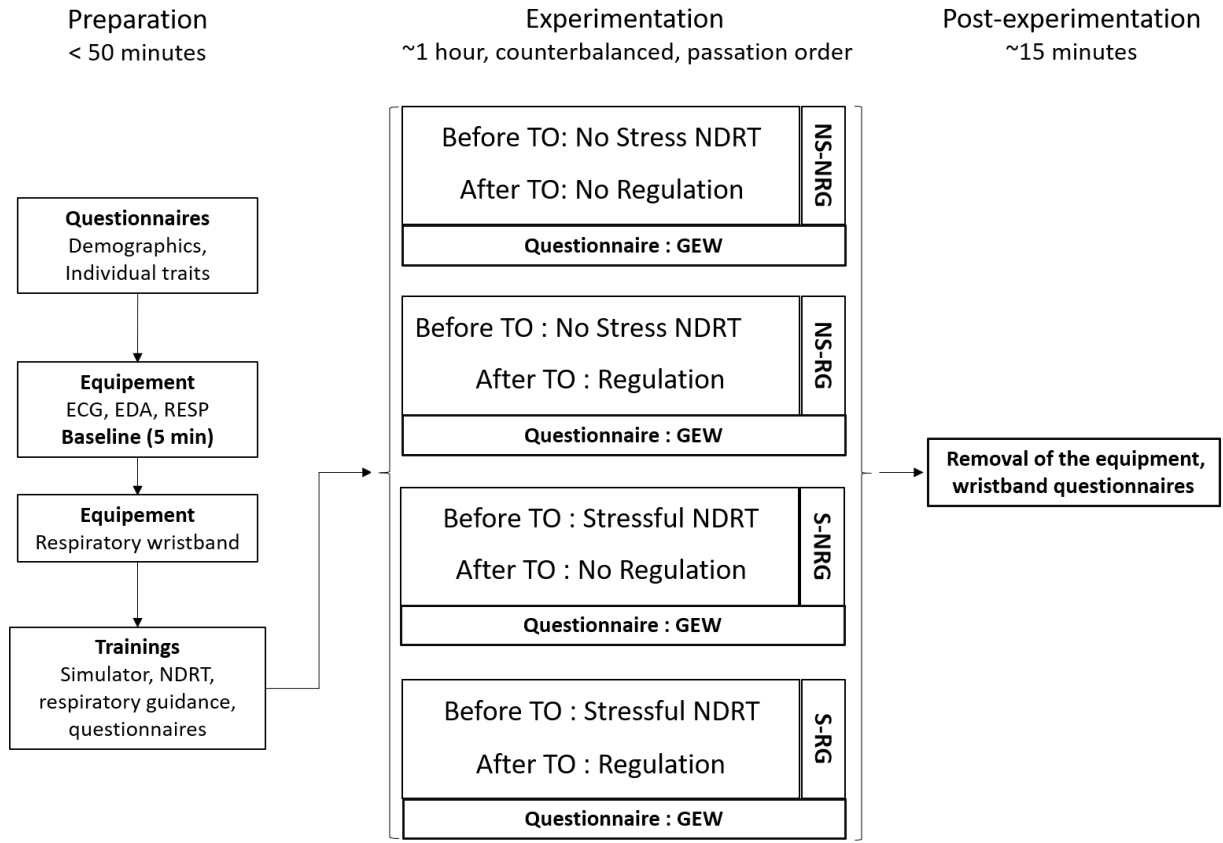

*Figure 5: Experimental course (TO = Takeover; GEW = Geneva Emotional Wheel; NDRT = Non Driving Related Task; NS = No Stress; S = Stress; NRG = No Respiratory Guidance; RG = Respiratory Guidance).*

To sum up, this is a 2x2 design, where the independent intrasujects variables were :

-Stress induction (2 levels) : *No stress NDRT vs Stressful NDRT.*

-Stress regulation (2 levels) : *No regulation vs Regulation.*

Leading to 4 experimental conditions, each of them presented to every participants :

-No stress, No Regulation (NS-NRG).

-No stress, Regulation (NS-RG).

-Stress, No Regulation (S-NRG).

-Stress, Regulation (S-RG).

**2.6 Statistics analysis**

ANOVA and Chi-square analyses were performed using JASP software (version 0.16), while logistic regression analyses were conducted using SPSS software (version 28.1).

### 2.6.1 Preliminary analysis regarding stress induction and respiratory guidance

Two-way ANOVAs with Stress (stress or No Stress) and Guidance (respiratory guidance or no respiratory guidance) were conducted to examine the main effects of stress induction and respiratory guidance on the data. McNemar's Chi-2 test was used to analyze the proportion of interaction with the NDRT after the takeover request. Cronbach's alpha was calculated for subjective questionnaires, indicating acceptable reliability ($\alpha > 0.7$; Taber, 2018). A significance threshold of $\alpha = 0.05$ was applied for all analyses. ANOVAs' effect sizes ($\eta^2$) were interpreted as small ($> 0.02$), medium ($> 0.13$), or large ($> 0.26$) (Bakeman, 2005).

After a preliminary verification of the stress induction, regarding the first objective of the study concerning the physiological efficacy of the respiratory guidance, participants' respiratory rate synchronization with the wristband's vibrations was analyzed, as well as HRV. The analysis focused on RR and RMSSD during the manual driving window and examined the main effect of regulation. In line with this objective, regarding the unobtrusiveness of the technique, driving safety during the use of respiratory guidance was assessed using driving parameters and usability evaluation from the participants. These parameters included mean speed and acceleration, SWRR, the standard deviation of the lateral position, lateral distance from overtaken vehicles, and the mean rating from the debrief questionnaire regarding the wristband.

### 2.6.2 Investigation of subjective regulation under individual dispositions

Regarding our second objective linked to the exploration of individual variations influencing the efficacy of the respiratory guidance, two groups were formed based on the variation of negative feelings between conditions S-NRG and S-RG. The "Regul" group (30 participants) experienced a reduction in negative feelings (variation: -0.2 to -2.6, M = -0.73, SD = 0.59), while the "NonRegul" group (34 participants) experienced an increase in negative feelings (variation: +0.2 to +1.8, M=0.66, SD=0.49). An independent sample t-test with a large effect size of d=2.58 showed a clear distinction between the two groups. Among the remaining participants, 13 participants had a null variation, and 6 had incomplete data.

The regressions were conducted in two steps and were conducted following recommendations from Brace et al. (2017). Step 1 consisted in three separate logistic regressions including questionnaires, physiological variables (measured during AD and MD in S-RG condition), and driving variables (measured during TO and MD in S-RG). Due to hands movements during driving, 62% of the EDA signals were not suitable to extract reliable EDA responses parameters, and consequently EDA data was not included in this regression analysis. Step 2 involved a regression combining the retained variables from Step 1. The stepwise descending method with the Wald test statistic was used for iterative variable elimination, with a Wald probability threshold of α=0.1. First iteration included only an intercept, and the $\chi^2$ goodness-of-fit test evaluated if uncertainty was significantly reduced when predictors were added in the following iterations. Model performance was assessed using composite tests of model coefficients in step I. In Step II, Cox & Snell's $R^2$ and Nagelkerke's $R^2$ values (explanatory power of the model), as well as the Hosmer-Lemeshow $\chi^2$ test (difference between predicted and observed groups), were considered. The significance of the retained variables in Step II was evaluated through the Wald test, along with their impact on the probability of belonging to the reference group (Exp (B) coefficients). The classification table indicated the percentage of correct classifications with a probability cutoff of 0.5 for assigning participants to the Regul group.

In addition, we also translated and reported some of the participants' feedback that could be useful to understand their impressions regarding the interaction with the proposed system.

## 3. Results

Detailed means for each variable in each condition can be found in the supplementary materials.

### 3.1 Stress induction

Significant main effects of stress were found for:

-Subjective evaluation (Figure 6) : Higher levels of negative feelings (F(1,76) = 39.54, p < .001, $\eta^2$ = .192) were reported when stress was present (M = 1.857, SD=0.1) compared to when stress was absent (M = 1.369, SD=0.1).

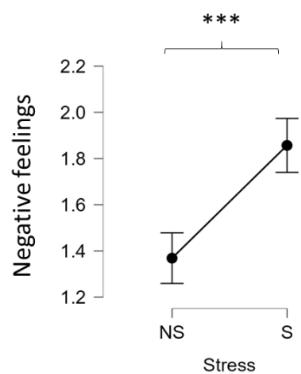

*Figure 6 : Negative feelings variation between the NoStress (NS) and Stress (S) conditions*

-Physiology (Figure 7): When stress was present, higher values were measured for HR ($F(1,77) = 82.69$, $p < .001$, $\eta^2 = .397$), RR ($F(1,78) = 31.55$, $p < .001$, $\eta^2 = .126$), sigh rate per minute ($F(1,80) = 11.23$, $p = .001$, $\eta^2 = .062$) and number of SCRs ($F(1,47) = 4.45$, $p = .04$, $\eta^2 = .032$).

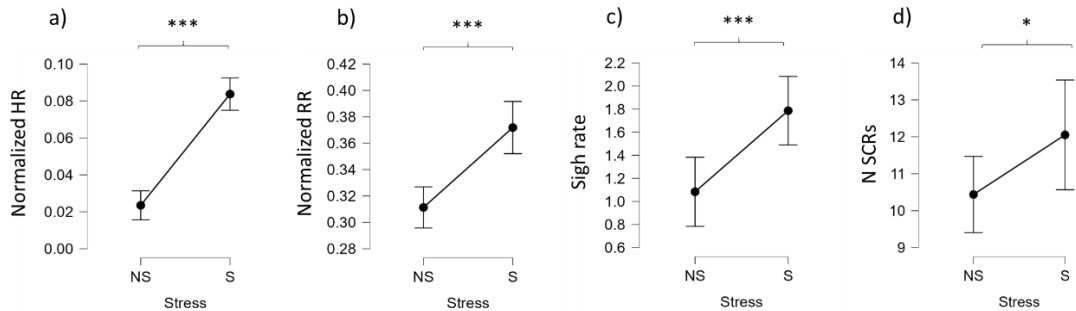

*Figure 7 : Physiological parameters variation between NoStress (NS) and Stress (S) conditions, for a) Heart Rate (HR); b) Respiratory Rate (RR); c) Sigh rate; d) Number of Skin Conductance Responses (N SCRs).*

-Behavior (Figure 8): The proportion of NDRT perseverance after the takeover request was higher when stress was present (45.2%) than absent (33.1%), indicating that participants tended to pursue their interaction with the NDRT despite the takeover request having been issued ($X^2(1, N = 153) = 4.87$, $p = .027$).

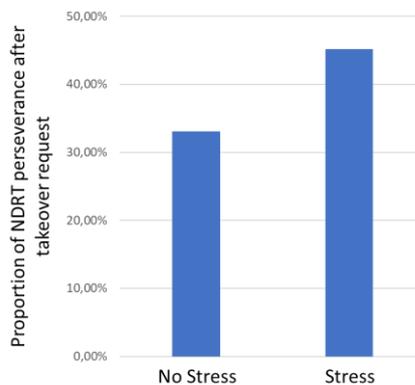

*Figure 8 : Proportion of NDRT perseverance after the takeover request*

### 3.2 Investigation of respiratory guidance

#### 3.2.1 Qualitative Feedback of Participants on the Respiratory Guidance

Regarding the participants' interaction with the wristband, we report the following feedback to capture their general impressions.

Some participants found that the wristband was easy to interact with and associated it with relaxation:

- "*The wristband did not distract me from driving; I had no trouble managing both*" (P27).
- "*I automatically synchronized with the vibrations without having to think about it*" (P39).
- "*The wristband allowed me to relax. It reminded me to breathe calmly*" (P52).
- "*I think my daily breathing is irregular. The wristband forced me to breathe more regularly, which was pleasant*" (P100).

Among the participants, it is interesting to note that some of them reported the relaxation effect was linked to the training phase and the difficulty of the game:

- "*I saw the wristband as relaxing; the training done before the start of the experiment allowed me to not see it as something new*" (P02).
- "*I really felt the relaxing effects of the wristband after the stressful game, and not when the game was neutral*" (P03).
- "*The wristband was rather neutral because my stress level was not high enough to see the relaxing benefit*" (P33).

Some participants reported more difficulty following the guidance while driving due to attentional demands:

- "*The biofeedback stressed me a bit because it required concentration; I think the same thing with sound would have been less stressful*" (P53).
- "*I think I have trouble concentrating on two things at once. The wristband rather stressed me*" (P35).
- "*I felt that the wristband was part of an 'all'; I experienced it as an additional constraint compared to driving*" (P37).
- "*I prioritized my concentration on driving when more complex actions like overtaking were required*" (P29).
- "*It was easy to interact with the wristband, but certain phases of driving made it more difficult to concentrate on it*" (P54).

Some participants reported that their driving experience influenced their interaction with the wristband:

- "*I am a good driver; it helped me manage the dual task with the wristband more easily*" (P13).

Two participants reported that synchronization with the wristband was difficult, notably due to the duration of the vibrations:

- "*The time between vibrations was insufficient for me. I had trouble coordinating with the wristband*" (P82).
- "*The vibrations were too fast*" (P62).

Some participants found that the experimental context forced them to follow the guidance:

- "*I experienced it as a constraint because I had to do it. In real life, I could use it more easily by having the choice*" (P84).

Finally, one participant reported that using sound instead of vibrations would have been easier:

- "*The wristband was okay, but I think the same thing with sound would be easier to relax with while driving*" (P28).

### 3.2.2 Effect of respiratory guidance on physiology

Regarding the respiratory rate, we found a significant effect of wristband activation ($F(1,70) = 284.93$, $p < .001$, $\eta^2 = .770$) during manual driving. The average RR was 18 cycles per minute in conditions without respiratory guidance, compared to 11 cycles per minute with respiratory guidance. The average reduction was 35% between the conditions with and without RG, which aligned with the target respiratory rate reduction. Figure 9 below shows the mean and the individual variation of respiratory rate between the conditions. Moreover, we found a significant augmentation of RMSSD with RG ($F(1,69) = 5.83$, $p < .05$, $\eta^2 = .036$).

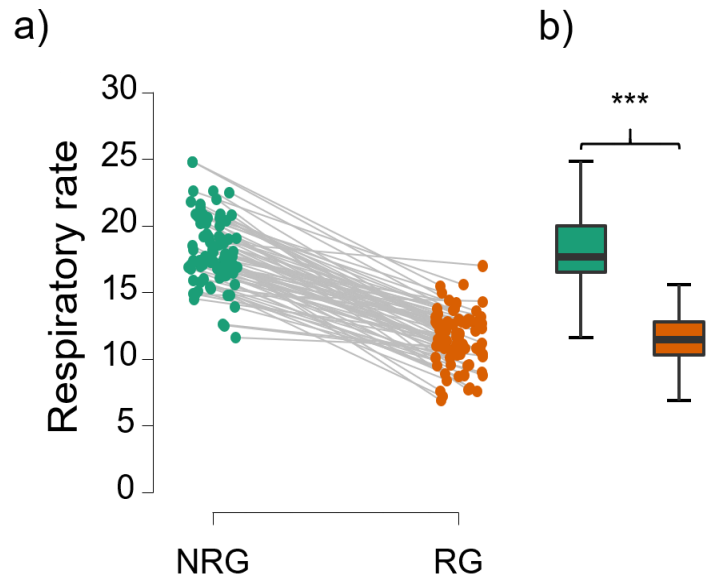

*Figure 9: Respiratory rate of each participant (a) and boxplot representation of the variation (b) between conditions with No Respiratory Guidance (NRG) and Respiratory Guidance (RG).*

### 3.2.3 Effect of the respiratory guidance on behavior and subjective questionnaires

Statistical analysis performed on the driving parameters did not show impacts of the respiratory guidance on the mean speed, acceleration, standard deviation of lateral position in the lane, lane departure rate, and distance from overtaken vehicles (p>.05). However, an effect of the regulation was evidenced regarding SWRR (F(1,74) = 21.951, p < .001, $\eta^2$ = .139), the SWRR was significantly larger when the wristband was not active (M = 6.52, SE = 0.37) than active (M = 5.39, SE = 0.27).

Regarding the subjective evaluation of the wristband using questionnaires, no effect was found on the negative feelings. The device evaluation ratings on a scale of 0 to 100 were as follows: the distraction caused by vibrations received a mean rating of 27.8 (SD = 24.9), while it was of 26.7 (SD = 24.3) regarding the difficulty of synchronizing respiration with vibrations, and of 15.3 (SD = 20.4) for the difficulty in distinguishing between inhalation and exhalation phases.

### 3.2.4 Individual variables impacting the efficacy of the wristband

We report below the results regarding the logistic regression conducted on parameters pre-selected in a first phase of analysis (see supplementary material). The following variables were studied:

-<u>Questionnaires</u>: Age group (19-35 years, 42-71 years); DBQ2 (inattention mistakes); DBQ6 (positive behaviors); Difficulty to discriminate inspirations/expirations phases from the wristband; Difficulty to synchronize with the wristband.

-<u>Driving variables (measured in S-RG)</u>: Takeover SWRR; Takeover mean speed; Lane departure rate in manual driving.

At the fourth iteration, the composite test of model coefficients is significant ($\chi^2$ (5)=27.260, p<.001). Cox and Snell's R² and Nagelkerke's R² are of 0.347 and 0.463, respectively. These values indicate that the model explains between 34% et 46% of the variance for the probability to belong in the Regul group. Hosmer-Lemeshow's test indicates that the values predicted by the model are not significantly different from observed values, indicating a good fit of the model to the data ($\chi^2$(8)=9.067, p > .1).

This model correctly classifies 76.6% of the observations (Table 1).

*Table 1: Classification table of the model.*

| Observed | Predicted | | |
|---|---|---|---|
| | NoRegul | Regul | % Correct |
| NoRegul | 27 | 7 | 79.4 (sensibility) |
| Regul | 8 | 22 | 73.3 (specificity) |
| Global correct % | | | 76.6 (accuracy) |

Table 2 compiles the variables included in the model and the associated statistics (Wald and exp beta values).

*Table 2: Summary of the variables retained in the model for classifying regulation.*

| Variable | B | E.S | Wald | ddl | Sig. | Exp(B) |
|---|---|---|---|---|---|---|
| DBQ$_{inatt}$ | 1.881 | 0.655 | 8.254 | 1 | 0.004 | 6.560 |
| DBQ$_{pos}$ | 1.595 | 0.584 | 7.466 | 1 | 0.006 | 4.926 |
| Difficulty to synchronize | -0.026 | 0.012 | 4.585 | 1 | 0.032 | 0.974 |
| SWRR takeover | -0.059 | 0.020 | 9.055 | 1 | 0.003 | 0.942 |
| Mean speed takeover | 0.094 | 0.035 | 7.054 | 1 | 0.008 | 1.099 |
| Intercept | -16.887 | 5.219 | 10.469 | 1 | 0.001 | 0.000 |

Regarding the odds ratio, a positive B indicates that an unit increase of the variable increases the likelihood of belonging to the Regul group by a factor of Exp(B). On the contrary, a negative B indicates that an increase of the variable decreases the likelihood of belonging to the Regul group by a factor of 1/Exp(B). The analysis of the odd ratios provides the following information:

- DBQ2 (Inattention errors): Individuals reporting driving errors due to inattention are more likely to be regulated by the wristband.
- DBQ6 (Positive behaviors): Individuals reporting positive driving behaviors are more likely to be regulated by the wristband.
- Difficulty to synchronize: Individuals experiencing difficulty in synchronizing with the wristband's vibrations are less likely to be regulated by it.
- SWRR takeover: A higher steering wheel reversal rate during the beginning of takeover phase in the S-RG condition reduces the chances of being regulated by the wristband.
- Mean speed takeover: A higher speed during the avoidance phase in the S-RG condition increases the chances of being regulated by the wristband.

## 4. Discussion

This study followed two objectives. The first objective was linked to the efficacy of the respiratory guidance in terms of unobtrusive respiratory rate reduction (that is, without manual driving perturbations). After checking that the protocol induced stress, and in regard to this first objective, parameters linked to respiratory rate and manual driving were analyzed. The second objective aimed at exploring the individual variations that may be linked to the subjective efficacy of the guidance in terms of stress regulation. Individual factors, behaviors, and perceptions were analysed in a iterative logistic regression process.

We discuss the results obtained in the following subsections, where "efficacy" refers to measurable performance, while "efficiency" relates to the effort required to achieve this performance (Eysenck et al., 2007).

### 4.1 Stress induction

In this study, stress was induced using a Non Driving Related Task (NDRT), that was completed during an autonomous driving phase. The use of this method allowed us to control and personalize the stress induction, but was also in line with the current development of autonomous driving, where new problematics, linked to the realization of NDRT, may arise due to the necessity to take back the manual control of the vehicle in specific occasions. Our study successfully induced stress among participants during automated driving using the NDRT. Arousal and negative feelings were increased in response to stress, as expected (Giannakakis et al., 2019; Hidalgo-Muñoz et al., 2019). Additionally, we found that the presence of stress resulted in increased perseveration towards the NDRT following the takeover alarm. This finding aligns with the attentional control theory proposed by Eysenck et al. (2007), suggesting an attentional bias towards stressors. In our experiment, this bias may have led to increased attention and difficulty in interrupting the NDRT. These first results show that participants were in a negative affective state prone to regulation, but also further underline the pertinence to regulate stress in such a context. However, despite our findings seems relevant in the context of automated driving, the fact that stress was not induced during the manual driving by ecological stressors (eg. High density traffic or weather) might limit the generalization of our conclusions regarding the use of regulation in manual driving.

**4.2 Physiological efficacy and unobstrusiveness of respiratory guidance**

Regarding respiratory guidance, an effective reduction in respiratory rate for all participants was noted. The average reduction achieved corresponded to the set rate for each participant (approximately 70% of the resting respiratory rate) and was accompagnied by an increased RMSSD, coherent with previous studies on slow breathing (Sevoz-Couche & Laborde, 2022; Zhou et al., 2023) . Regarding manual driving, results showed an impact of the wristband activation on the SWRR. As previous studies showed that this parameter is sensitive to dual tasks in driving (Kountouriotis et al., 2016) it may reflect a distraction from the driving elicited by the wristband. However, the effect size of the variation was medium, and no other impact on critical driving parameters, such as speed and lane position was evidenced.

Regarding the evaluation of the wristband, ratings collected with the subjective questionnaires tend to illustrate that participants did not evaluate the wristband as being neither distracting nor difficult to use. However, no global impact on the subjective regulation evaluated via the emotional wheel was evidenced.

Based on the debriefs, it is interesting to point out some discrepancies amongst participants on their impressions of the interaction. Specifically, the influence of the habituation phase and of the perceived contextual demand (regarding the driving task, but also regarding the game), on the usability of the wristband was noted, but also the influence of anterior driving experience. The use of sound instead of vibrations was proposed by one participant, echoing the conclusions from Lee, Elhaouij and Picard (2021). Interestingly, some participants noted that the experimental context itself had an influence on their comfort regarding the use of the wristband. This can be linked with the well-known discussion regarding the impact of manipulation checks on the data collected (Hauser et al., 2018), pointing out the importance to carefully design ecological experimentations on regulation technologies. Finally, some participants reported that, despite a careful personalization, the rhythm of the vibrations was too fast. We believe that these debriefs allows to further points out the importance of the personalization of regulation systems, and to suggest that further studies on regulation allows the participant to adjust himself some parameters. Complementary to these feedbacks, the subsequent quantitative analysis allowed a deeper understanding and discussion of the technology used.

### 4.3 Influence of individual disposition on the subjective efficacy of respiratory guidance

In line with the second objective, two balanced and heterogeneous groups were formed based on variations in subjective feelings between conditions with and without respiratory guidance. The groups consisted of individuals who experienced either a decrease or an increase in negative feelings with the respiratory guidance. Individual dispositions influencing the belonging to either group were studied using logistic regressions.

#### 4.3.1 Prosocial driving habits

Our results showed an influence of prosocial driving habits on subjective regulation. This variable has been shown to reflect empathic abilities (Bainbridge et al., 2022; Karras et al., 2022), which are strongly correlated with self-awareness and bodily states awareness (Ernst et al., 2013; Preston et al., 2007). The contribution of prosocial behaviors may be interpreted as being related to the facility to perceive and appraise their own physiological and mental state as relaxed due to slow breathing. However, this interpretation remains speculative as we did not measure certain variables that could support or refute it. Indeed, while we considered interoception through the use of THISQ (Vlemincx et al., 2021), a more specific measurement of interoceptive awareness may bring more insights.

#### 4.3.2 Inattention errors driving habits

Regarding the influence of inattention errors habits, the literature links this dimension to the establishment of automatized behaviors in driving (Guého et al., 2014). We can interpret the influence of this variable in relation to the functioning of the wristband, which seems to require a certain level of attention and conscious modulation of breathing to be properly followed. Although RR of all participants decreased and driving was successfully managed with the guidance (i.e., equivalent efficacy), participants with higher levels of driving attentiveness may have required additional resources to follow the wristband (i.e., reduced efficiency). This reduction in efficiency would have increased the need of participants to divide their attentional resources between driving and the wristband, leading to discomfort and stress. On the other hand, participants with lesser driving attentiveness would have had a greater attentional availability. This would have led to a greater efficiency in focusing on the wristband. This conclusion is also supported by the participant's feedback, were some of them indicated that they had difficulties to focus on both driving and the wristband.

These interpretations seems to be in line with previous results obtained using a slightly different technology : Béquet et al., 2021 used tactile stimulations to mimic a decreased cardiac rhythm. They showed that the efficacy of the technic to reduce stress was directly linked to the attentional focus of the participant towards the stressor itself. We hypothesize that, in the current study, the increase in attentional resources allocated to the wristband could have contributed to reducing the stressful impact of the occupational stress generated by the NDRT combined with the takeover and with the driving situation itself. In this regard, in addition to its physiological impacts, the wristband could also have acted as an attentional deployement strategy amongst these participants.

### 4.3.3 Perception of the wristband

Results indicate that the evaluation of the interaction with the wristband also influenced the efficacy of subjective regulation. A higher perceived difficulty in synchronizing with the wristband was associated with a lower probability of belonging to the group of regulated participants. Consistent with our previous interpretations, this difficulty may be linked to a greater attentional demand to achieve comparable efficacy, influencing the overall subjective experience.

It should be noted that we did not find a correlation between inattention errors in driving (declared by DBQ) and the perception of synchronization difficulty. This suggest that these two variables made distinct contributions to the probability of being regulated, despite being both linked to a difference in resource mobilization efficiency. The increased demand to follow the respiratory guide in some participants might be both explained by cognitive (e.g. working memory) or interoceptive/somesthetic sensitivity factors. In light of this interpretation, specific measurements via classic tasks (eg., heartbeat counting task regarding interoceptive sensibility, Schandry, 1981) would have been relevant to consider.

### 4.3.4 Driving parameters

Finally, we found an influence of driving parameters measured during the takeover. A higher SWRR decreased the probability of being regulated, while an increase in speed increased this probability. Higher SWRR could reflect an increased focus of participants on driving at the beginning of the takeover (Kountouriotis et al., 2016). Regarding speed, previous work established that its increase reflects a reduced perception of the criticality of a given situation (Deniel, 2019). Overall, less attentive participants with a lower perception of the takeover's criticality showed enhanced stress regulation during manual driving. This suggests that the guidance may be less effective if the context of use of the guidance is perceived as more critical or attention-demanding, which is coherent regarding our interpretations in previous sections.

### 4.4 Design recommendations

Based on the results regarding the factors influencing the breathing guidance, the following recommendation regarding the design and implementation of such regulation in a driving context can be made:

**Customization of vibration patterns**: The rhythm of the wristband's vibrations should be adjusted to align with slower respiratory rates. The study found that the mean delivered rhythm was 9.7 BPM, but slower respiratory rates have been shown to have more beneficial effects (Zaccaro et al., 2018). Moreover, some participants found the vibrations too fast and had trouble coordinating with the wristband. Therefore, providing customizable settings for vibration intervals that allow users to select or automatically adapt to slower rhythms might help users better align their breathing with the guidance and enhance the relaxation effects.

**Attention to attentional resources**: The wristband's design should minimize the attentional resources required for detection and synchronization with the guidance during driving. This can be achieved by customizing the vibration force and ensuring the design facilitates easy attentional focus. Allowing for a longer familiarization period can help users automate the use of the wristband, reducing the cognitive load during actual use. Additionally, delivering isolated stimulations to induce a sigh, which has psychophysiological benefits (Vlemnincx et al., 2013), rather than continuous stimulations that demand constant attention, would be relevant to implement and test.

**Contextual consideration**: The difficulty and criticality of the usage context should be carefully considered as they may impact participants' efficiency in following the guidance. For instance, some participants reported adjusting their guidance depending on the complexity of the driving task. Thus, is would be relevant to implement an adjustment of the guidance's intensity based on real-time assessment of driving complexity. Such solutions are known as "contextualy aware interventions" (eg. Balters et al., 2019).

**Driving experience consideration**: In the same vein, participants with higher driving experience found it easier to manage the dual task of driving and following the wristband. Therefore, the system could include a calibration phase that assesses driving experience and adjusts the complexity of the breathing guidance accordingly. The exploration of other modalities of regulation for drivers that need a greater attentional focus on driving would be interesting to conduct in further studies.

**Flexible Use Conditions**: Some participants felt constrained by the experimental context, suggesting that real-life application should allow for more flexibility. The design should enable users to engage with the wristband at their own pace and discretion, enhancing its practical usability without feeling forced.

While we test one haptic breathing guidance, it would be ideal to tailor regulation systems to the participants, developing solutions based either on various haptics localization (eg. Miri et al., 2020; Umair et al., 2021), or various sensory preferences, such as sound feedback as pointed out by one participant and by previous studies (eg. Zhou et al., 2024; Zepf et al., 2020; Lee et al., 2021).

The use of wristbands for guided breathing, however, brings several relevant applications : First, it present the advantage of being private, as pointed out by (Miri et al., 2020). It can also be used in various contexts, other than driving, and the hardware supporting the intervention, a wristband, is very common on the market (as Zhou et al. 2023 demonstrated using a dedicated app running on an Apple Watch). Finally, the use of a wristband system allows to embed wearable stress measurement technologies, using EDA or photoplethysmography signals that enable to trigger "just-in-time" automatic interventions (Zepf et al. 2020).

### 4.5 Limitations

This study presents certain limitations influencing its generalizability. First, since this study was conducted during the COVID-19 pandemic, participants had to wear masks. This prevented us from controlling whether inspiration was consistently done through the nose, as it could impact the efficacy of regulation (Zaccaro et al., 2018). Second, regarding subjective data collection, the reliability of questionnaires depends on participants' introspection abilities and of potential biases such as desirability bias, expectancy bias, or respondent fatigue (Alekhine et al., 2020). Regarding the selected questionnaires for this study, while their inclusion was based on their relevance and reliability, it is worth noting that alternative questionnaires could have been considered as well.

### 5. Conclusions

The results obtained in this study, on a qualitative and quantitative standpoint, and their interpretations align with prior studies (Bequet et al., 2022; Xu et al., 2021; Miri et al., 2020, Zhou et al., 2023) and emphasize the intricate relationship between perceived attentional demand and context in determining the efficiency and efficacy of respiratory guidance for stress reduction. In the context of driving, our findings reveal the variability of this demand based on individual factors related to driving styles, appraisal of both context criticality and of the technique itself. Based on these results, we offered several recommendations and perspectives for the design of unobtrusive stress regulation methods in driving.

Finally, it is important to note that despite the promising insights from our study, our model only accounts for approximately 34% to 46% of the variance in the probabilities of belonging to the regulated group. It is likely that unconsidered factors, such as tactile sensitivity and perception of context criticality influence the effectiveness of using the guidance. Future studies should explore and specify these factors and consider the use of various machine learning algorithms for automated classification (e.g., Evin et al., 2022).

To sum-up, this study brings the following contributions :

First, this experiment was the first (to our knowledge) to test the effect of a wristband device to regulate stress in the context of the takeover of an autonomous vehicle. Regarding this technology, the present study also allowed the replication of results obtained by previous studies such as the one from Zhou et al. (2023), and brought additional results regarding the individual characteristics to considers when using such device.

Second, our study points out the criticality to consider the notions of efficiency vs efficacy of tactile respiratory technics in an applied ergonomics standpoint.

Finaly, this study's results showed the importance to consider carefully the dynamic between individual characteristics and the driving context itself. These results allowed to provide design recommendations regarding the implementation of breathing regulation, and to propose research perspective in this field.

## 6. Acknowledgments

The authors would like to thank Bruno Piechnik for his technical assistance regarding the NDRT, Bertrand Richard for his insights regarding script development, and Myriam Evennou for her help with the statistical analysis. This work was funded by University Gustave Eiffel.

UNOBTRUSIVE DRIVING RESPIRATION GUIDANCE